# Generalized image deconvolution by exploiting spatially variant point spread functions

SangYun Lee[1,2], Kyeoreh Lee[1,2], Seungwoo Shin[1,2] and YongKeun Park[1, 2,3*]
[1] Department of Physics, Korea Advanced Institute of Science and Technology (KAIST), Daejeon 34141, Republic of Korea.
[2] KAIST Institute Health Science and Technology, Daejeon 34141, South Korea
[3] Tomocube Inc., Daejeon 34051, Republic of Korea.
*Correspondence:: yk.park@kaist.ac.kr

**An optical imaging system forms an object image by recollecting light scattered by the object. However, intact optical information of the object delivered through the imaging system is deteriorated by imperfect optical elements and unwanted defects. Image deconvolution, also known as inverse filtering, has been widely exploited as a recovery technique because of its practical feasibility, and operates by assuming the linear shift-invariant property of the imaging system. However, shift invariance is not rigorously hold in all imaging situations and it is not a necessary condition for solving the inverse problem of light propagation. Here, we present a method to solve the linear inverse problem of coherent light propagation without assuming shift invariance. Full characterization of imaging capability of the system is achieved by successively recording optical responses, using various laser illumination angles which are systematically controlled by a digital micro-mirror device. Experimental results show that image distortions caused by optical defocus can be restored by conventional deconvolution, but severe aberrations produced by a tilted lens or an inserted disordered layer can be corrected only by the proposed generalized image deconvolution. This work generalizes the theory of optical imaging and deconvolution, and enables distortion-free imaging under any general imaging condition.**

An optical imaging system delivers information about an object by collecting scattered light from the object and projecting it onto an imaging plane. In theory, the light scattered from each point is treated as a new point source, and forms a one-to-one correspondence with other points on the imaging plane. Ideally, the matrix of connected in-and-output spatial frequency components of an ideal imaging system can be represented as an identity matrix (Fig. 1a). In a practical imaging system, however, the image of a point object never forms a point image. This is because of the limited spatial bandwidth of the optical imaging system, which is determined by a finite numerical aperture (NA) and aberrations.

    To compensate for aberrations, deconvolution (or inverse filtering), which has remarkable practical feasibility, has been widely used [1,2,3]. Deconvolution is based on two assumptions: the linearity and the shift invariant properties. Linearity in imaging implies that if the output responses to the input point sources are known, then the object image will be expressed as a linear combination of the output responses corresponding to the linear constituents of the input information. Shift invariance means that the translational shifts of a source in the object plane do not change the profile of the output response (Fig. 1b). In coherent and incoherent imaging systems, the linearity and shift invariant properties are applied to complex optical fields, and to light intensity profiles, respectively. When those assumptions hold, the output distribution of the complex amplitude (or intensity) at the imaging plane can be expressed as the input complex amplitude (or intensity) convolved with a unique point spread function (PSF) of the imaging system. In other words, it is assumed that one PSF works over the entire field of view. In the Fourier domain of a coherent imaging system, under this assumption, the optical spectrum of the complex amplitude distribution at the imaging plane is simply a multiplication of the object optical spectrum and the Fourier transform of the PSF, or a coherent transfer function (CTF).

    Accordingly, the spatial frequency of an object is transferred to the imaging plane independently of other spatial frequencies, and the matrix in Fig. 1b connecting the input and output frequency components of the linear shift invariant imaging system becomes a diagonal matrix. Deconvolution theory has been applied to a wide range of optical imaging systems, from telescope imaging[4], positron-emission tomography[5] to deconvolution microscopy[6], confocal microscopy[7,8], and digital

holography[9,10]. It has also been reported that if *a priori* knowledge about an object is assumed to be known, the resolution of the recovered images can be further improved[11].

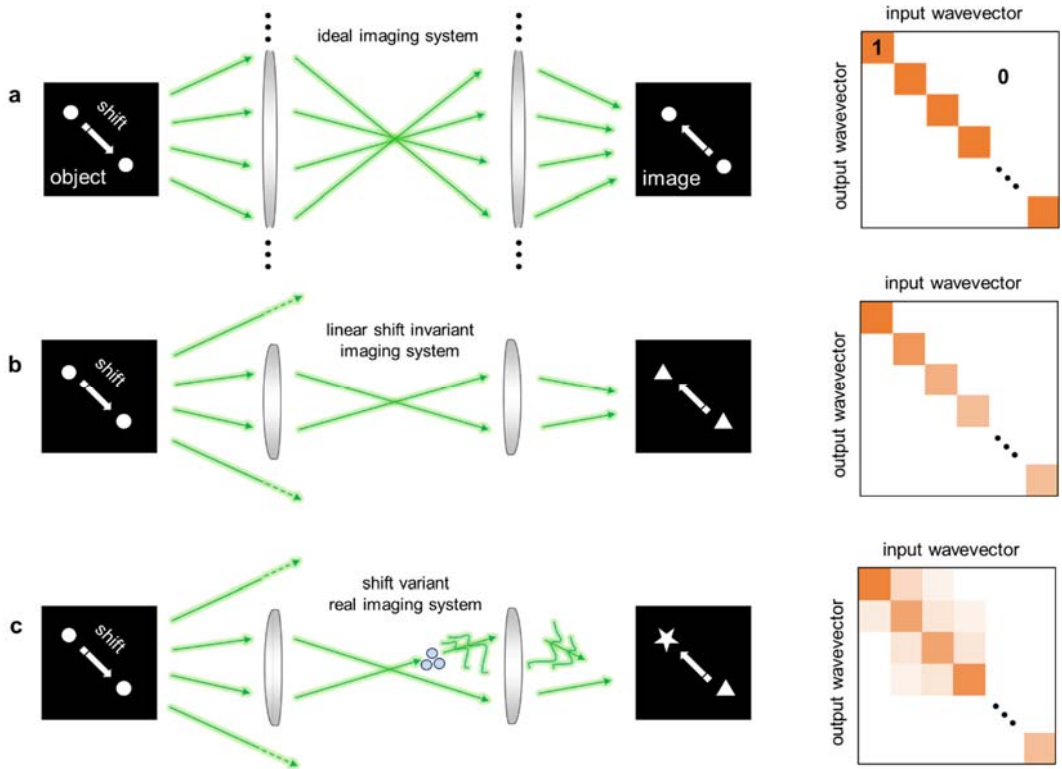

Figure 1 | A schematic of the image forming capability of (a) ideal, (b) linear shift invariant, and (c) linear shift variant imaging systems.

Although deconvolution has been used as a standard method for correcting aberrations, the shift invariance is approximately valid only when moderate aberrations exist. For instance, misalignment in an optical imaging system or contamination by dust can cause the PSF (or CTF) of an imaging system to be different at each position of the FOV, because the separated point sources in the object plane have spatially distinct optic paths. Similarly, one of the main issues in adaptive optics[12-15] for deep tissue imaging is to properly correct for spatially varying aberrations imposed by complex tissue structures. Recently, a wavefront shaping technique by utilizing multi-conjugate planes[16,17] or turbid layer conjugation[18] has been proposed, but is practically limited to fully account the mixing of light spatial modes in an imaging plane due to a finite size of both a pixel and a total area of current spatial light modulators (SLMs). Even in an aligned imaging system, which maintains a high degree of shift invariance by keeping the FOV much smaller than the size of the optical elements, it is inevitable that the matrix in Fig. 1c characterizing the image delivery capability in the frequency domain has some non-diagonal components. These non-diagonal components can be an important issue, especially when a high level of aberration correction is required. Therefore, in order to completely compensate for aberrations imposed by imperfections in the imaging system, the PSF of a system which varies depending on the position of the object in the FOV, need to be systematically considered.

Here, we generalize the concept of image deconvolution by introducing the concept of variant PSFs. By characterizing all the responses of an imaging system to the translational shift of an input point source over the entire FOV with the help of wavefront shaping techniques[19-24], the full characterization of the image forming capability of the imaging system is accomplished. Inspired by recent advances in techniques[25-32] for characterizing light propagation in scattering media, our approach measures the transmission matrix (TM) of a coherent imaging system in the Fourier domain.

To investigate the formation of optical imaging – one of the fundamental principles in photonics - we revisit the inverse problem of light propagation through a coherent imaging system[33] to obtain distortion-free object images without assuming shift invariance. We used an interferometric microscope equipped with a digital micro-mirror device (DMD), which enables the full-field measurements of an optical field, with the capability of controlling the wavefront of an incident beam onto a sample[34] (see Methods).

Using this system, the propagation of light or the TM information of the used imaging system was characterized by successively recording the transmitted light fields for various illumination angles of an incident beam that was controlled by a DMD. Then, to demonstrate the principle of the generalized image deconvolution, optical fields with both distorted amplitude and phase objects under various aberrations were corrected, using two approaches, the conventional and the generalized image deconvolutions, respectively. We demonstrate that the present approach effectively corrects system aberrations in nearly all imaging situations, whereas the conventional deconvolution only works when the linear shift invariance of an imaging system holds.

**Principle**

The full characterization of an imaging system was achieved by measuring the TM of the system. The optical responses of the imaging system were successively recorded to a set of impulses spanning the frequency space. Here, we displayed a set of input binary patterns generated by the Lee hologram method[34,35] as impulses on a DMD, and recorded the holograms of the transmitted beam with a Mach-Zehnder interferometric microscope. However, it should be noted that the formulas that follow are applicable to any type of SLMs, including a liquid crystal on silicon and a galvanometer mirror, and any input basis that is capable of generating object space.

Suppose a total number of $N$ input fields, spanning the spatial frequency space, are generated within a chosen NA. Let it be called a scan NA. Then, an input frequency spectrum of the $i_{th}$ binary image $\mathbf{y}_i$ can be expressed as $\mathbf{y}_i = \sum_{j=1}^{\alpha} d_{ji} \mathbf{k}_j^{in}$, $i=1…N$, where $\{\mathbf{k}_j^{in}\}_{j=1…\alpha}$ forms a basis of the DMD frequency space up to the NA of the condenser lens, and the coefficient $d_{ji}$ is directly determined by taking the Fourier transform of the $i_{th}$ binary image. Likewise, the output frequency spectrum of an image captured by the camera which corresponds to the $i_{th}$ binary input becomes $\mathbf{z}_i = \sum_{j=1}^{\beta} c_{ji} \mathbf{k}_j^{out}$, $i=1…N$. In this case, $\{\mathbf{k}_j^{out}\}_{j=1…\beta}$ expands the spatial frequency space of the camera plane up to the NA of the imaging objective lens. Let D and C be matrices of the input and output optical spectra whose elements are $\{d_{ij}\}_{i=1…\alpha, j=1…N}$ and $\{c_{ij}\}_{i=1…\beta, j=1…N}$, respectively. The TM of an imaging system T connects the spatial frequency components of the input optical spectrum in the $\mathbf{k}_j^{in}$ basis to ones of the output spectrum in the $\mathbf{k}_j^{out}$ basis as follows: $\mathbf{z}_i = \mathbf{T} \mathbf{y}_i$. Inserting the expansions of the $\mathbf{y}_i$ and $\mathbf{z}_i$ in $\mathbf{k}_j^{in}$ and $\mathbf{k}_j^{out}$ bases into the last equation, we obtain $c_{\gamma i} = \sum_{j=1}^{\alpha} t_{\gamma j} d_{ji}$, where $t_{\gamma j}$ is the $(\gamma, j)$-entry of T for $\gamma = 1…\beta$. Multiplying $d_{i\delta}^{-1}$, the $(i, \delta)$-entry of $\mathbf{D}^{-1}$, to both sides and summing over the index $i$ finally gives

$$\mathbf{T} = \mathbf{C}\mathbf{D}^{-1}. \tag{1}$$

In Fig. 2b, the square matrix parts of the C and D matrices are represented. The calculated TM is physically the same as having the information about all the CTFs (or PSFs) at each point of the FOV. Thus, the CTF at each point in the FOV of an imaging system can be retrieved from the measured TM. We retrieved the representative CTF of the imaging system by propagating a point image source at the center of the FOV, which corresponds to the scan NA, through the TM (Fig. 2c). If we expand the CTF, $\mathbf{O}$, using the $\{\mathbf{k}_j^{out}\}$ basis as $\mathbf{O} = \sum_{j=1}^{\beta} o_j \mathbf{k}_j^{out}$, the coefficient $o_j$ can be mathematically written as,

$$o_j = \sum_{i=1}^{\alpha} t_{ji} \zeta_i, \text{ where } \zeta_i = 1 \text{ if } |k_i^{in}| \leq \text{scan NA,}$$

(2) (Fig. 2c)

$$\zeta_i = 0 \text{ otherwise}$$

Unless otherwise stated in this paper, the representative CTF refers to the CTF that the imaging system has at the center of the FOV.

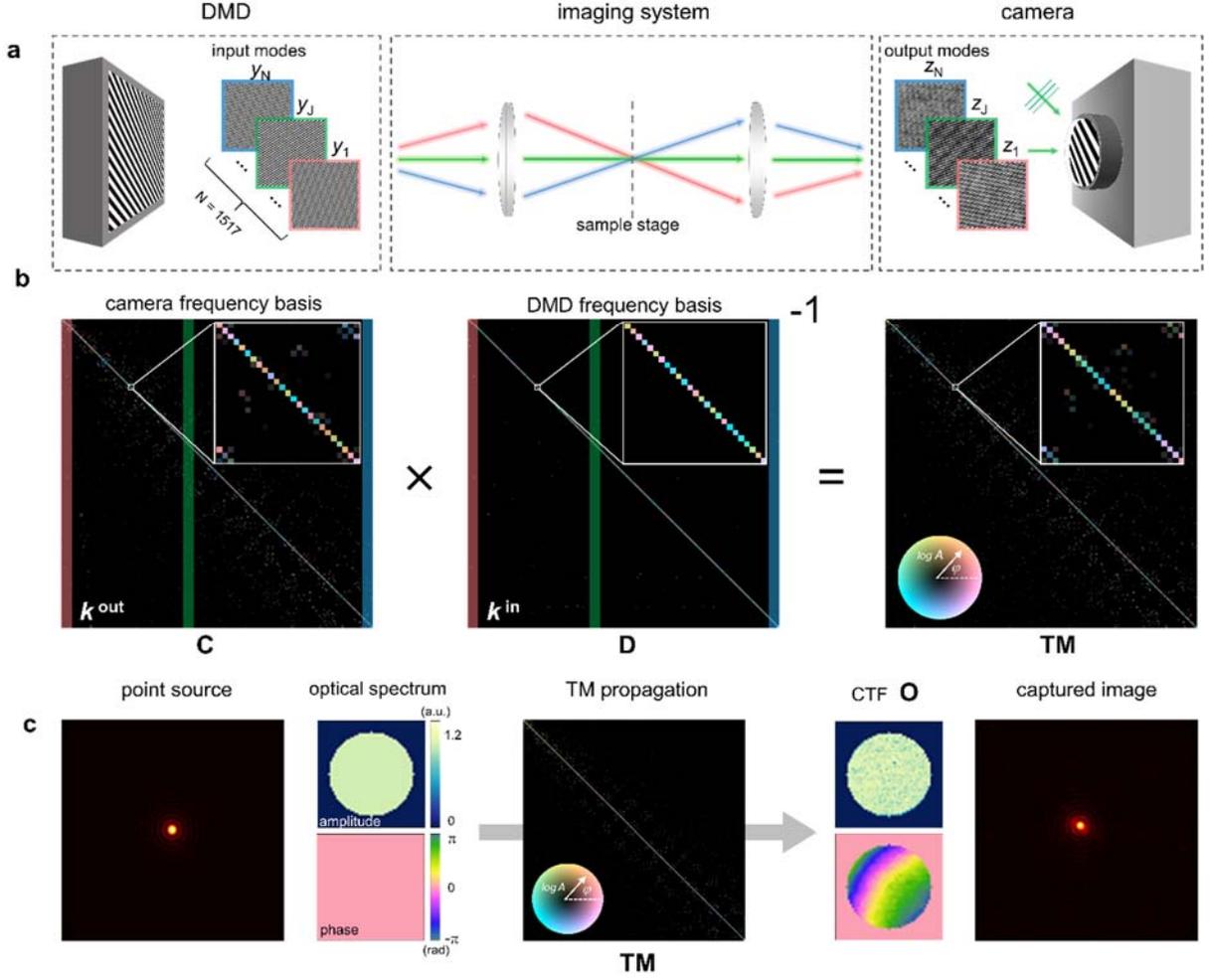

Figure 2 | Characterization of imaging capability of the optics system (a) Generated binary input DMD patterns and the recorded output interferograms after passing through the imaging system. (b) The characterization process of the TM. (c) The retrieval procedure for a CTF.

By fully characterizing an imaging system by measuring the TM, any distortion in the imaging can be corrected by using a back propagation of the complex optical fields. From the measured hologram of a sample, the complex optical fields can be retrieved using the proper field retrieval algorithms[36] (Fig. 3a). Let the optical spectrum of the sample image S be expressed as $\mathbf{S} = \sum_{j=1}^{\infty} s_j \mathbf{k}_j^{out}$. Then, in the conventional deconvolution approach, the distortion-free optical spectrum of the sample S′ can be obtained by dividing S by O from Eq. (2) as,

$$\mathbf{S}' = \sum_{j=1}^{\beta} s_j' \mathbf{k}_j^{out}, \text{ where } s_j' = \frac{s_j}{o_j} \text{ if } \left|\mathbf{k}_j^{out}\right| \leq \text{scan NA} \quad (3)$$

(Fig. 3b)

$$s_j' = s_j \text{ otherwise.}$$

In contrast, in the proposed approach, the generally deconvoluted optical spectrum of the sample S″ can be retrieved from S by applying the inverse of TM, $\mathbf{T}^{-1}$, from Eq. (1) as,

$$\mathbf{S}'' = \mathbf{T}^{-1}\mathbf{S} = \mathbf{D}\mathbf{C}^{-1}\mathbf{S}.$$

(4) (Fig. 3c)

The remaining practical task for retrieving S″ is to obtain an inverse of C. This can be numerically achieved when C decomposes into $\mathbf{U\Sigma V}^{\dagger}$ via the single value decomposition, where $\mathbf{U}$ and $\mathbf{V}$ are complex unitary matrices and $\mathbf{\Sigma}$ is a diagonal matrix. Accordingly, $\mathbf{C}^{-1}$ can be directly obtained by calculating $\mathbf{V}(1/\mathbf{\Sigma})\mathbf{U}^{\dagger}$. Taking the inverse Fourier transform of S′ and S″ then finally gives the conventionally and generally deconvoluted sample images, respectively (Figs. 3b and 3c).

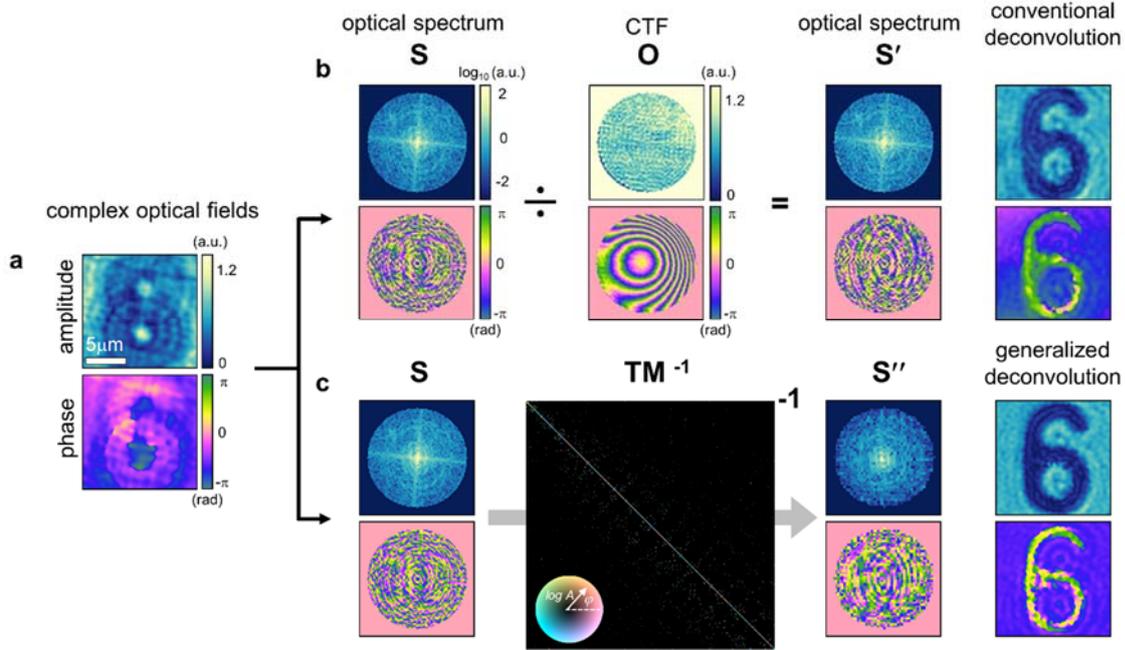

Figure 3 | Correction method for existing aberrations in the imaging system (a) Retrieved 2-D complex optical fields of a raw hologram. (b) The conventional image deconvolution process. (c) The generalized image deconvolution process. For simplicity, only the optical spectrum within the scan NA is displayed.

## Experimental setup

To illuminate light fields with desired spatial frequencies on the object plane and to record the optical responses of the imaging system, we employed Mach-Zehnder light interferometry[34] (Fig. 4a). The characterization of the imaging system, i.e., TM measurement, was achieved by recording a series of holograms corresponding to input fields. A DMD was used to sequentially generate the light fields with specific spatial frequencies. The maximum FOV and the resolution of the imaging system are in principle determined by the resolution and maximum frequency of the spatial frequency space being spanned, which is controlled by the DMD. The TM of the imaging system is measured before loading a sample, and then the optical field of a sample under a vertical laser incidence is measured.

To validate the image transfer capability in the presence of various aberrations, the alignment of the Mach-Zehnder interferometer was intentionally changed, as shown in Fig. 4b. In detail, optically defocused, lens-tilted, and scattered systems were established by slightly elevating the imaging objective lens, tilting the aspheric tube lens of the 4-*f* imaging configuration from the aligned positions, and inserting disordered lens papers into the optic axis, respectively (see Methods).

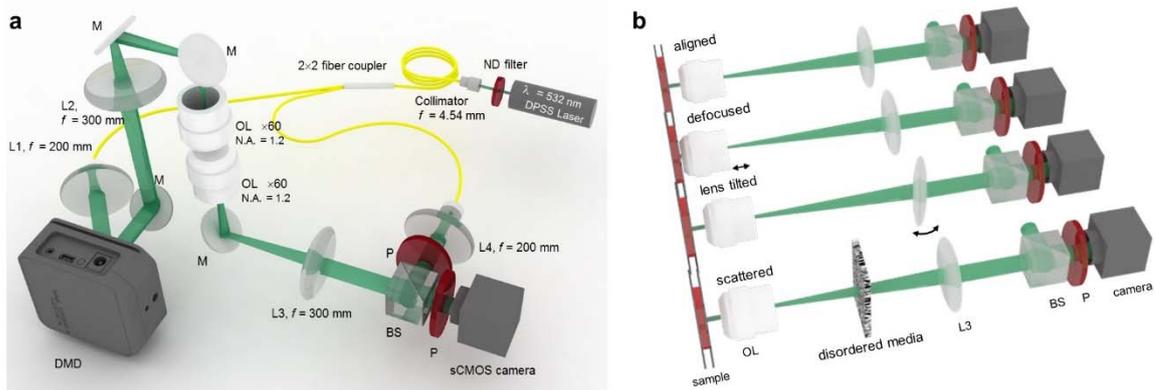

Figure 4 | The experimental optics setup (a) An inverted microscope based on Mach-Zehnder interferometry. A sample is loaded between the condenser and imaging objective lenses. L: lens; M: mirror; OL: objective lens; BS: beam splitter; P: polarizer (b) Imposition of various aberration conditions on the imaging system, including optical defocus and lens tilt, and insertion of weakly scattering media in the optic axis.

## Result

Violations in the linear shift invariant property of an imaging system

In order to demonstrate how imposed aberrations on the imaging system deteriorate the condition of shift invariance, two simulated point sources, whose optical spectrums are uniform up to NA of 0.6 (Fig. 5a), were propagated through the measured TMs under various optically misaligned conditions. If the imaging system has the linear shift invariant property, then the PSF and the CTF will not be changed by laterally translating an input point source. Therefore, the degree of shift invariance of the imaging system can be mapped over the entire FOV by calculating the correlations with the CTF or the profile of the PSF corresponding to the center of the FOV.

As an example, the CTFs of the lens-tilted imaging system at three different positions (α, β, and O in Fig. 5a) are shown in Fig. 5b. The two-dimensional (2-D) correlation map of the PSFs over the entire FOV of the lens-tilted system, representing the degree of shift invariance of the system, is presented in Fig. 5c. Following the same procedure, the 2-D PSF correlation map of the aligned, optically defocused, lens tilted, and scattered systems are given in Figs. 5d–g, respectively. The output complex optical fields of the two point sources (α and β in Fig. 5a) through the measured TMs, were also presented right to the correlation map of the corresponding imaging system.

The correlation map of the aligned optics setup (Fig. 5d) shows a high degree of the shift invariant property. The output amplitude and phase profiles of the propagated two point sources are also similar to those of the input light field (Fig. 5a), except for minor distortions originating from unintentional misalignments and defects in the optical elements.

In the case of the optically defocused system (Fig. 5e), the imaging system still exhibits a high degree of shift invariance. This can be explained by the nature of light propagation. Optical defocus, or light propagation in general, only creates phase shifts on the Fourier optical spectra, so the generation of new spatial wave vectors from other spatial frequencies does not occur. Accordingly, in principle, short translations of a lens along the optic axis do not impose violations of shift invariance of the imaging system. In the propagated output field, two point foci no longer exist due to dissatisfaction of the imaging condition; only weak interference patterns can be observed (Fig. 5e). In the phase map, however, two developed concentric phase rings are observed and this is consistent with the theory of light propagations.

In the case of tilting the tube lens (Fig. 5f), however, it severely violates the shift invariance of the imaging system. This is because optical responses upon translational shifts of an input source in the object plane are not isotropic in the imaging plane anymore. Instead, the shift invariant property of the system is only preserved along the particular direction which corresponds to the axis of the lens rotation. It is also observed that the output complex optical fields of the two point sources are squeezed along the corresponding direction.

When a disordered layer is inserted in the optic axis, which diffuses an incident plane wave into multispectral random waves, it effectively abolishes the shift invariant property of the aligned imaging system (Fig. 5g). As a result, the 2-D PSF correlation map of the scattered imaging system has a remarkably narrow range in which the shift invariance is preserved and this narrow region at the FOV center is where the widely-known optical memory effect[37-40] comes in. Also, when the point light sources are propagated through the measured TM, the output complex field is diffused out around the original point locations.

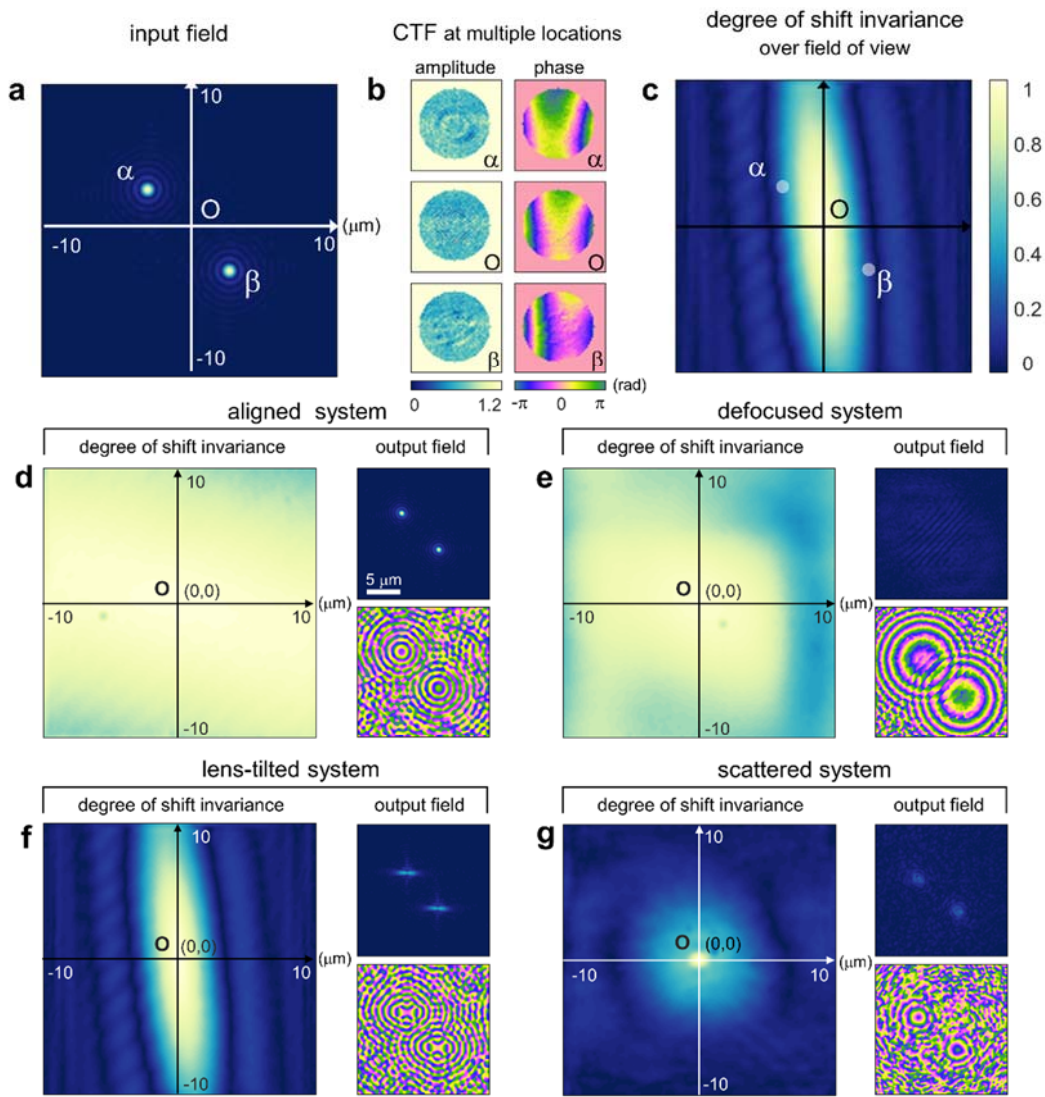

Figure 5 | Degree of shift invariant property of the imaging systems with imposed aberrations (a) The input complex optical field of two simulated point sources. (b) The CTFs of the lens-tilted imaging system at the center of the FOV, and at the positions of the two point sources. (c) The 2-D correlation map of the PSF profiles over the entire FOV of the lens-tilted imaging system, indicating the degree of shift invariance of the system. 2-D PSF correlation maps and the output complex optical fields of the two point sources of the (d) aligned, (e) defocused, (f) lens tilted, and (g) scattered imaging systems.

Reconstruction of the aberration-free complex optical field image
In order to demonstrate the capability of distortion-free imaging using the generalized image deconvolution techniques, we corrected the optical field images of objects measured under optically aligned and misaligned conditions. A 1951 USAF target and a polystyrene bead with a diameter of 10 μm were used as an amplitude object and a phase object, respectively. A set of binary patterns was generated using the DMD to achieve both the maximum spatial frequency and the FOV up to 0.92 $\mu m^{-1}$ and $20 \times 20\ \mu m^2$, respectively (see Methods and Materials).

The retrieved complex optical fields of the samples under various imaging conditions are presented in Figure 6. The measured representative CTFs of the aligned (Fig. 6a), optically defocused (Fig. 6b), lens tilted (Fig. 6c), and scattered (Fig. 6d) imaging systems are given. Restored images obtained with both the conventional deconvolution and the proposed generalized deconvolution are presented. To quantitatively analyze the quality of the image reconstruction, the correlation value of each sample field with the generally deconvoluted field in the aligned imaging system is indicated.

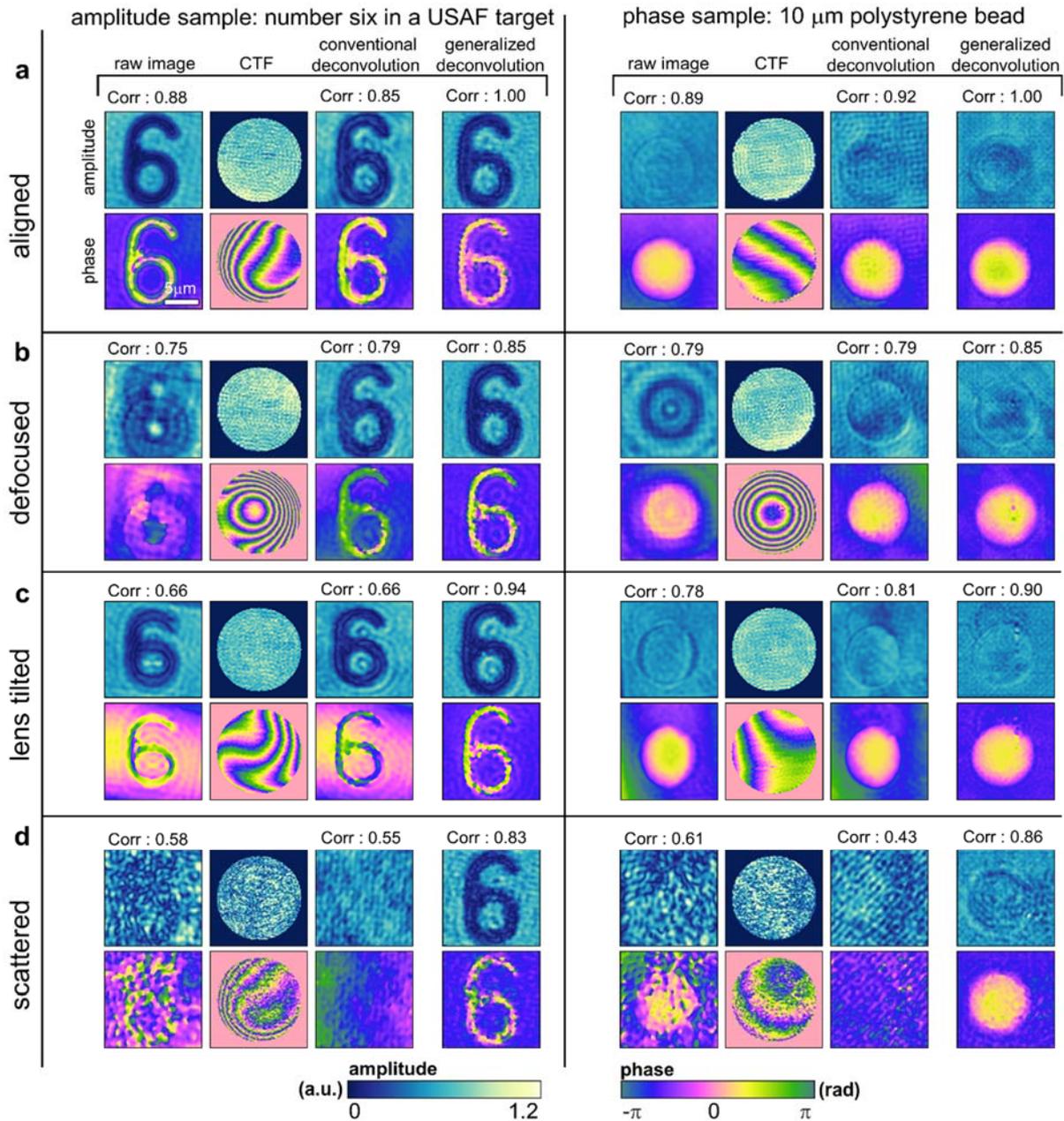

Figure 6 | Reconstruction of distortion-free optical fields of an amplitude and a phase target. Retrieved optical field of a USAF target and 10-μm-diameter polystyrene bead from the raw holograms, the restored optical fields of the targets using both conventional and generalized deconvolutions, and the CTF of the optical configuration in (a) aligned, (b) defocused, (c) lens tilted, and (d) scattered imaging conditions. For each complex optical field image, the value of the field-field correlation with the optically aligned condition is also presented.

The quality of image reconstruction using the present generalized deconvolution method is significantly enhanced in comparison to the conventional devolution method. The conventionally deconvoluted optical images exhibit the correct profile of the object only when the imaging system is aligned (Fig. 6a) or optically defocused (Fig. 6b). In contrast, the restored optical fields obtained with the generalized deconvolution give the correct shapes of the object even in the lens-tilted (Fig. 6c) and scattered (Fig. 6d) systems, notwithstanding some speckle noises.

In the optically aligned imaging system (Fig. 6a), the profiles of both conventionally and generally deconvoluted complex optical fields are similar to those of the original holograms, as expected. While conventional deconvolution preserves the non-uniform background phases of the original hologram, the generalized deconvolution further corrects the background phase profile so that it is uniform. This aspect can be also found in the optically defocused and lens-tilted imaging systems. Consequently, the restored complex optical fields based on the generalized image deconvolution exhibit a flat background phase, even in the scattered imaging system.

Next, in the optically defocused system (Fig. 6b), the corrected profiles of the samples again validate both the conventional and generalized image deconvolution approaches for correcting aberrations. The concentric phase rings of the CTFs clearly show the well-known phase profile of a spherical wavefront kernel.

In the tube-lens-tilted imaging system (Fig. 6c), however, we can see that only the generalized deconvolution properly restored the fine shapes of the number six and the spherical bead from the distorted optical fields. Despite some de-blurring effects, conventional deconvolution could not reshape the distorted images of either the USAF target or the polystyrene bead, because a tilted lens in the imaging system violates the shift invariance of the imaging system.

Imperfections in the distortion correcting capability of the conventional deconvolution seem more apparent when it comes to the scattered imaging system, which mimics an extreme situation that occurs with many microscopic scatters in the optic beam path (Fig. 6d). In this case, the application of conventional deconvolution to the raw holograms even strengthens the image distortions. The corresponding CTF of the scattered system also exhibits speckle-like inhomogeneous amplitude and phase profiles. The generalized image deconvolution, however, still restores the shape of both the amplitude and phase objects with moderate spatial noises.

Conclusion

In this paper, we introduced a generalized image deconvolution concept which does not require the assumption of a spatially invariant system, and experimentally demonstrated its image reconstruction performance. To accomplish this, the light propagating behaviors of the imaging system were fully characterized by successively measuring the optical responses of the system to a set of illumination beams. Our results show that the assumption of spatial invariance is satisfied only for some ideal cases. The present method correctly restored the complex amplitude images of both amplitude and the phase objects despite various types of aberrations, including a tilted lens and the insertion of a scattering layer. In contrast, the conventional deconvolution method could only correct image distortions when the distortions did not violate the shift invariance, as in the case of optical defocus.

The present work demonstrates the proof-of-principle of the suggested idea for a coherent imaging system. However, the present approach is general and also readily applicable to any imaging configuration including incoherent, x-ray, terahertz, and infrared imaging systems. For an incoherent system such as fluorescence microscopy, the intensity measurements of the points sources at various locations within the FOV can be used.

The 2-D PSF correlation maps in Fig. 5 representing the degree of shift invariance of the imaging system provide a neat comparison between the conventional and generalized image deconvolutions. The main advantage of conventional image deconvolution is that image restoration is simple and fast based on the invariant CTF (or OTF) of the imaging system. Meanwhile, generalized image deconvolution can compensate any complex aberrations by eliminating the assumption of shift invariance in the imaging system, but complicated calculations are inevitable since all changes in the CTF based on the positions within the FOV must be considered.

From a practical point of view, if the FOV of an imaging system can be divided into small sections where the shift invariance holds locally, even complicated aberrations which violate the global shift invariance of the system can also be effectively corrected, by assigning the proper CTFs to the corresponding subsections. The most efficient way to compensate for aberrations would be to minimize the number of small sections that maintain the shift invariance locally and constitute the entire FOV of the imaging system.

In sum, considering that the shift invariance is always deteriorated to some extent, even in optically aligned systems, we expect that this generalized imaging deconvolution method can be effectively exploited, particularly in industrial fields and biomedical applications which require a high level of aberration corrections for the purpose of inspecting optical elements and high-performance telescopes, and calibrating microscopes.

**Methods and Materials**

Optics setup

To demonstrate the aberration correcting capability of the generalized image deconvolution, we employed Mach-Zehnder interferometry (Fig. 2a). A diode-pumped solid state laser with a single longitudinal mode ($\lambda$ = 532 nm, 50 mW, Cobolt Co., Solna, Sweden) served as the light source. A 2×2 fiber optic coupler ($\lambda$ = 532 ± 15 nm, 90:10 split, Thorlabs, U.S.A.) connected to a fiber collimator ($f$ = 4.34 mm, NA = 0.57, Thorlabs, U.S.A.) divides the incident laser beam into a sample and a reference arm. The illumination angles of the laser beams impinging onto the sample are systematically controlled by using a DMD (V-9501, Texas Instruments Inc., U.S.A.), which is located in the conjugate plane of the sample. Two identical objective lenses (UPlanSApo, 60×, water immersion, NA = 1.2, infinity corrected, Olympus Inc., U.S.A.) were used as a condenser and an imaging objective lens, respectively. However, when imaging a thick object, an objective lens with a long working distance (LUCPlanFL N, 60×, WD = 1.7 mm, NA = 0.7, Olympus Inc., U.S.A.) was used as a condenser lens. Total magnification of the optical system is 100×, contributed by a telescopic 4-$f$ system which consists of a 60× imaging objective lens and an aspheric lens with a focal length of 300 mm. The sample and the reference arm were then recombined by a 50:50 beam splitter in front of the high-speed sCMOS camera (528×512 pixels with a pixel size of 6.5 μm, Neo sCMOS, ANDOR Inc., Northern Ireland, U.K.), and two polarizers were used to adjust the intensities of the sample and the reference beam intensities to maximize the fringe visibility.

Various aberrations imposed on the optical imaging system

To impose various aberrations on the Mach-Zehnder setup, the optical components were intentionally misaligned via image defocus, lens tilt, and insertion of weakly scattering media in the optic axis, as shown in Fig 2b. Defocus was achieved by slightly moving the imaging objective lens from a focused position. To apply more severe aberrations, the aspheric tube lens of the 4-$f$ imaging system in front of the beam splitter was tilted by 10 degrees. To generate an extreme situation, we inserted two pieces of lens papers between the imaging and pupil planes.

Transmission matrix recording

To record responses of the optical imaging system for various light illumination angles, i.e., measuring the TM, we constructed a signal triggering circuit which connects both the DMD and the sCMOS camera. The DMD controls the angle of light illumination and then a trigger signal is sent to the camera, so that the corresponding interferogram is captured before the next DMD pattern is projected.

The control of illumination angles was achieved by projecting Lee holograms onto the DMD[35]. Detailed information can be found elsewhere[34]. To measure the TM, the angular frequency space of the imaging system was sequentially scanned. The total number of DMD patterns is determined by both the scanned NA and the step size in angular frequency space. Because inverse filtering is a band-limited technique, the scan NA is fundamentally limited by the value of min($NA_c$, $NA_i$), where $NA_c$ and $NA_i$ denote the NAs of the condenser and imaging objective lenses, respectively. Meanwhile, the step size determines the maximum FOV of objective images to be corrected without aliasing artifact. The speed of the TM recording process is determined by the frame rate of the camera.

In the experiments with a 10-μm-dimeter polystyrene bead and a 1951 USAF target, we chose the scan NA = 0.6 and FOV = 20×20 μm$^2$. Accordingly, the scanning step size was set to be 1/20 μm$^{-1}$ and the total step numbers were then 1,517. The frame rate of the camera was 67 Hz. The entire TM measurement takes less than 30 sec. To correct temporal phase noise during the TM measurements, a reference pattern is displayed on the corner of the DMD.

References:

1. Frieden, B. R. in *Picture Processing and Digital Filtering*     177-248 (Springer, 1975).
2. Stroke, G. & Zech, R. A posteriori image-correcting "deconvolution" by holographic fourier-transform division. *Physics Letters A* **25**, 89-90 (1967).
3. Castleman, K. Digital image processing.   (1993).
4. Starck, J., Pantin, E. & Murtagh, F. Deconvolution in astronomy: A review. *Publications of the Astronomical Society of the Pacific* **114**, 1051 (2002).
5. Colsher, J. G. Fully-three-dimensional positron emission tomography. *Physics in medicine and biology* **25**, 103 (1980).
6. McNally, J. G., Karpova, T., Cooper, J. & Conchello, J. A. Three-dimensional imaging by deconvolution microscopy. *Methods* **19**, 373-385 (1999).
7. Shaw, P. J. & Rawlins, D. J. The point-spread function of a confocal microscope: its measurement and use in deconvolution of 3-D data. *Journal of Microscopy* **163**, 151-165 (1991).
8. Sarder, P. & Nehorai, A. Deconvolution methods for 3-D fluorescence microscopy images. *Ieee Signal Proc Mag* **23**, 32-45 (2006).
9. Cotte, Y., Toy, M. F., Pavillon, N. & Depeursinge, C. Microscopy image resolution improvement by deconvolution of complex fields. *Optics Express* **18**, 19462-19478 (2010).
10. Cotte, Y. *et al.* Marker-free phase nanoscopy. *Nature Photonics* **7**, 113-117 (2013).
11. Frieden, B. R. Restoring with maximum likelihood and maximum entropy. *JOSA* **62**, 511-518 (1972).
12. Booth, M. J. Adaptive optics in microscopy. *Philosophical Transactions of the Royal Society of London A: Mathematical, Physical and Engineering Sciences* **365**, 2829-2843 (2007).
13. Kubby, J. A. *Adaptive Optics for Biological Imaging*.   (CRC press, 2013).
14. Girkin, J. M., Poland, S. & Wright, A. J. Adaptive optics for deeper imaging of biological samples. *Current opinion in biotechnology* **20**, 106-110 (2009).
15. Booth, M. J. Adaptive optical microscopy: the ongoing quest for a perfect image. *Light: Science & Applications* **3**, e165 (2014).
16. Simmonds, R. D. & Booth, M. J. Modelling of multi-conjugate adaptive optics for spatially variant aberrations in microscopy. *Journal of Optics* **15**, 094010 (2013).
17. Wu, T.-w. & Cui, M. Numerical study of multi-conjugate large area wavefront correction for deep tissue microscopy. *Optics express* **23**, 7463-7470 (2015).
18. Park, J.-H., Sun, W. & Cui, M. High-resolution in vivo imaging of mouse brain through the intact skull. *Proceedings of the National Academy of Sciences* **112**, 9236-9241 (2015).
19. Vellekoop, I. M. & Mosk, A. Focusing coherent light through opaque strongly scattering media. *Optics Letters* **32**, 2309-2311 (2007).
20. Park, J.-H. *et al.* Subwavelength light focusing using random nanoparticles. *Nature photonics* **7**, 454-458 (2013).
21. Yu, H. *et al.* Recent advances in wavefront shaping techniques for biomedical applications. *Current Applied Physics* **15**, 632-641 (2015).
22. Yoon, J., Lee, K., Park, J. & Park, Y. Measuring optical transmission matrices by wavefront shaping. *Optics Express* **23**, 10158-10167 (2015).
23. Rotter, S. & Gigan, S. Light fields in complex media: mesoscopic scattering meets wave control. *arXiv preprint arXiv:1702.05395* (2017).
24. Yoon, J. *et al.* Optogenetic control of cell signaling pathway through scattering skull using wavefront shaping. *Scientific Reports* **5** (2015).
25. Popoff, S. *et al.* Measuring the transmission matrix in optics: an approach to the study and control of light propagation in disordered media. *Physical review letters* **104**, 100601 (2010).
26. Yu, H. *et al.* Measuring large optical transmission matrices of disordered media. *Physical review letters* **111**, 153902 (2013).
27. Popoff, S., Lerosey, G., Fink, M., Boccara, A. C. & Gigan, S. Controlling light through optical disordered media: transmission matrix approach. *New J Phys* **13**, 123021 (2011).
28. Choi, Y. *et al.* Overcoming the diffraction limit using multiple light scattering in a highly disordered medium. *Physical review letters* **107**, 023902 (2011).

29   Mosk, A. P., Lagendijk, A., Lerosey, G. & Fink, M. Controlling waves in space and time for imaging and focusing in complex media. *Nature photonics* **6**, 283-292 (2012).
30   Tao, X., Bodington, D., Reinig, M. & Kubby, J. High-speed scanning interferometric focusing by fast measurement of binary transmission matrix for channel demixing. *Optics Express* **23**, 14168-14187, doi:10.1364/OE.23.014168 (2015).
31   Boniface, A., Mounaix, M., Blochet, B., Piestun, R. & Gigan, S. Transmission-matrix-based point-spread-function engineering through a complex medium. *Optica* **4**, 54-59 (2017).
32   Sarma, R., Yamilov, A. & Cao, H. Enhancing light transmission through a disordered waveguide with inhomogeneous scattering and loss. *Applied Physics Letters* **110**, 021103 (2017).
33   Kim, T., Zhou, R., Goddard, L. L. & Popescu, G. Solving inverse scattering problems in biological samples by quantitative phase imaging. *Laser & Photonics Reviews* **10**, 13-39 (2016).
34   Shin, S., Kim, K., Yoon, J. & Park, Y. Active illumination using a digital micromirror device for quantitative phase imaging. *Optics Letters* **40**, 5407-5410 (2015).
35   Lee, W.-H. Binary computer-generated holograms. *Applied Optics* **18**, 3661-3669 (1979).
36   Debnath, S. K. & Park, Y. Real-time quantitative phase imaging with a spatial phase-shifting algorithm. *Optics Letters* **36**, 4677-4679 (2011).
37   Feng, S., Kane, C., Lee, P. A. & Stone, A. D. Correlations and fluctuations of coherent wave transmission through disordered media. *Physical review letters* **61**, 834 (1988).
38   Bertolotti, J. *et al.* Non-invasive imaging through opaque scattering layers. *Nature* **491**, 232-234 (2012).
39   Katz, O., Small, E. & Silberberg, Y. Looking around corners and through thin turbid layers in real time with scattered incoherent light. *Nature photonics* **6**, 549-553 (2012).
40   Katz, O., Heidmann, P., Fink, M. & Gigan, S. Non-invasive single-shot imaging through scattering layers and around corners via speckle correlations. *Nature Photonics* **8**, 784-790 (2014).

Supplementary information accompanies this paper at

Acknowledgements
This work was supported by KAIST, BK21+ program, Tomocube, and the National Research Foundation of Korea (2015R1A3A2066550, 2014M3C1A3052567, 2014K1A3A1A09063027).

Author Contributions
Y.P. conceived and supervised the project. S.L. performed the experiments and analyzed the data. K.L. and S.S. provided analysis tools. All authors wrote the manuscript.

Author Information Reprints and permissions information is available online at www.nature.com/reprints. Correspondence and requests for materials should be addressed to Y.P. (yk.park@kaist.ac.kr).

Competing financial interests
The authors declare no competing financial interests.